\documentclass[prl,10pt,twocolumn,
 amsmath,amssymb,
 aps,
]{revtex4-1}

\usepackage{todonotes}
\usepackage{graphicx}
\usepackage{dcolumn}
\usepackage{subfigure}
\usepackage[normalem]{ulem}
\usepackage{bm}

\usepackage{todonotes}

\newcommand{\twoone}{(1,2,1)}
\newcommand{\twotwo}{(1,2,2)}
\newcommand{\bmnot}{($l$,$m $,$n$)}

\begin{document}

\title{Superwalking droplets}

\author{Rahil N. Valani$^{1}$}\email{rahil.valani@monash.edu}
\author{Anja C. Slim$^{2,3}$}
\author{Tapio Simula$^{1,4}$}
\affiliation{$^1$School of Physics and Astronomy, Monash University, Victoria 3800, Australia} 
\affiliation{$^2$School of Mathematical Sciences, Monash University, Victoria 3800, Australia}
\affiliation{\mbox{$^3$School of Earth, Atmosphere and Environment, Monash University, Victoria 3800, Australia}}
\affiliation{$^4$Centre for Quantum and Optical Science, Swinburne University of Technology, Melbourne 3122, Australia}

\begin{abstract}
A \textit{walker} is a droplet of liquid that self-propels on the free surface of an oscillating bath of the same liquid through feedback between the droplet and its wave field. We have studied walking droplets in the presence of two driving frequencies and have observed a new class of walking droplets, which we coin \textit{superwalkers}. Superwalkers may be more than double the size of the largest walkers, may travel at more than triple the speed of the fastest ones, and enable a plethora of novel multi-droplet behaviors.
\end{abstract}

\maketitle

A liquid bath oscillating vertically may sustain bouncing droplets on its free surface \citep{Walker1978}. For sufficiently large accelerations of the bath, these bouncing droplets may begin to walk \citep{Couder2005}. Such walkers emerge just below the Faraday instability threshold, above which the liquid-air interface of the bath becomes unstable and standing subharmonic Faraday waves emerge \cite{Faraday1831a}. In the walking state, the droplet is propelled by local Faraday waves that it excites on each bounce and whose decay time is a function of the proximity to the Faraday threshold, also referred to as the \textit{memory} of the system. At high memory, waves generated by the droplet in the past continue to affect the motion of the droplet. Interactions of a walker with barriers and other droplets are mediated by this wave field, resulting in rich dynamics and a variety of multi-droplet configurations \cite{Pucci2016Non-specularDroplets,Protiere2006,Borghesi2014,PhysRevFluids.3.013604,PhysRevE.78.036204,PhysRevFluids.2.053601,Lieber2007,Eddielattice,Eddi2008,ValaniHOM,twodroplets}. This hydrodynamic system comprising a wave and a droplet mimics several features of the quantum realm including orbital quantization in rotating frames \citep{Oza2014} and harmonic potentials \citep{Perrard2014a,Perrard2014b}, wave-like statistical behavior in confined geometries \citep{PhysRevE.88.011001,Sáenz2017}, tunneling across submerged barriers \citep{Eddi2009}, and has been predicted to show anomalous two-droplet correlations \cite{ValaniHOM}. A detailed review is provided by Bush \cite{Bush2015}. 

\begin{figure}
\centering
\includegraphics[width=\columnwidth]{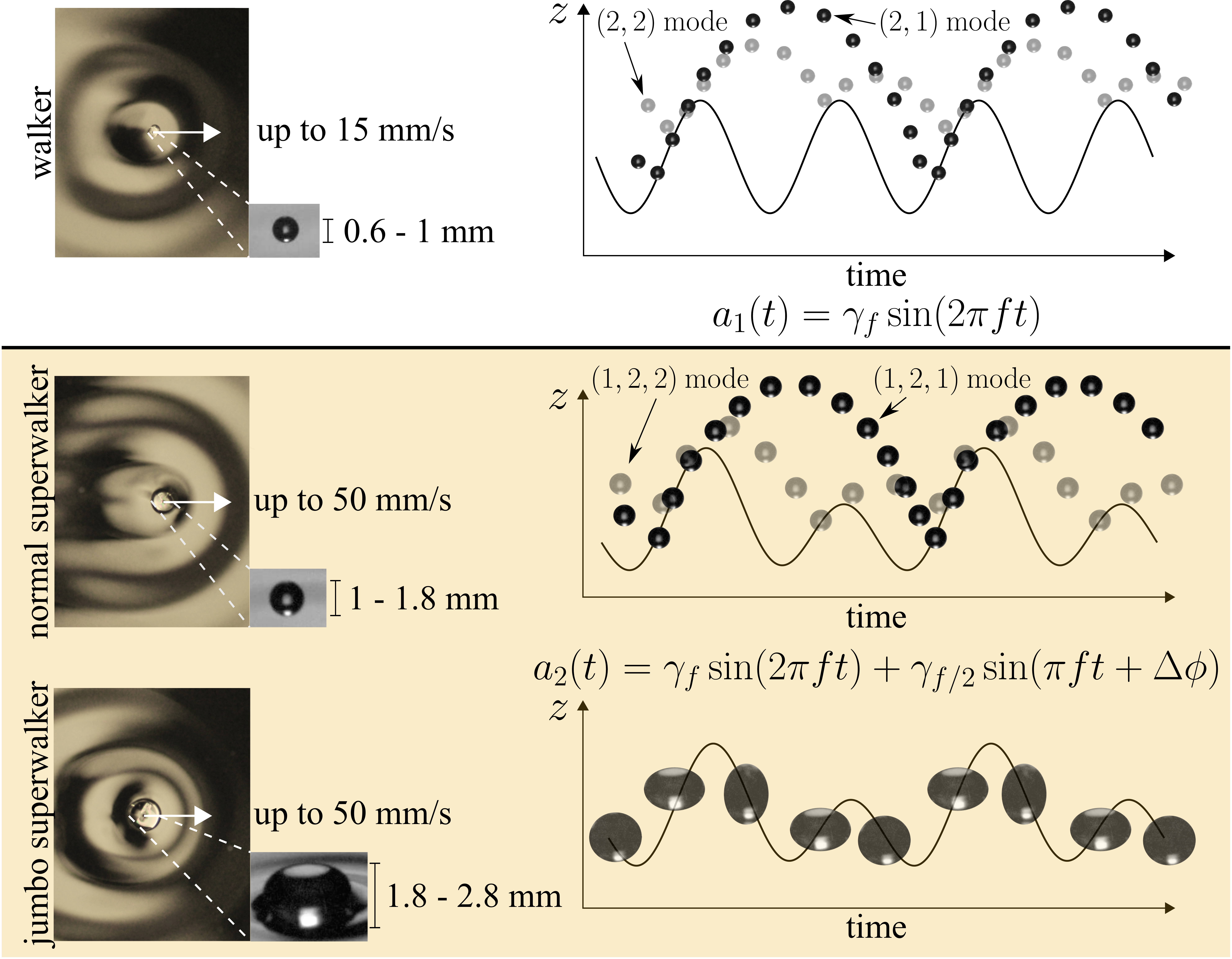}
\caption{Comparison of a walker (top), a normal superwalker (middle) and a jumbo superwalker (bottom). Superwalkers emerge when the bath is driven at two frequencies $f$ and $f/2$ with a phase difference $\Delta\phi$. They may be significantly larger than walkers and may move significantly faster. Left panels show top views of typical droplets and their wave fields, and side views of the same droplets.  Right panels show the bath motion (solid curve) and the typical bouncing motion of the droplets, see text for notation.}
\label{Fig: schematic}
\end{figure}

Walkers are typically driven by a pure sine wave  $a_{1}(t)=\gamma \sin(2 \pi ft)$, where $\gamma$ is the amplitude of the driving acceleration, $f$ is the driving frequency, and $t$ is time. For a commonly studied system of silicone oil with $20\,\text{cSt}$ viscosity
and $f=80\,\text{Hz}$, droplet radii of $0.3\,\text{mm}$ to $0.5\,\text{mm}$ and walking speeds of up to $15\,\text{mm}/\text{s}$ have been observed \citep{Molacek2013DropsTheory,Wind-Willassen2013ExoticDroplets}. Here we present a new class of walking droplets, which we coin \textit{superwalkers}, that emerge when the fluid bath is driven simultaneously at a frequency $f$ and the subharmonic frequency $f/2$ with a phase difference $\Delta\phi$ according to the acceleration
\begin{equation} \label{eq:acceleration}
a_2(t)=\gamma_{f} \sin(2 \pi f t)+\gamma_{f/2} \sin( \pi f t+\Delta\phi).
\end{equation}
In the commonly studied system noted above, superwalkers can be significantly larger than walkers with radii up to $1.4\,\text{mm}$ and they can walk at up to $50\,\text{mm}/\text{s}$. The largest superwalkers undergo significant internal deformation and barely lift off from the surface of the bath. We call these \textit{jumbo superwalkers}. The key differences between a walker and the two kinds of superwalkers are summarized in the schematic of Fig.~\ref{Fig: schematic}. Fundamental differences between walkers and superwalkers are also evident in their inter-droplet interactions. Due to their large inertia, superwalkers may easily overcome the wave barrier that typically prevents contact interactions between walking droplets, enabling superwalkers to form a variety of novel stationary and dynamic bound states.  

Our experiments were performed using a circular bath of diameter $18\,\text{cm}$ filled to a height of approximately $8\,\text{mm}$ with silicone oil of viscosity $20\,\text{cSt}$. The bath was mounted on the cone of a subwoofer speaker \citep{Harris2016}, which was placed on an optical breadboard. The quality of uniaxial vibrations and leveling of the bath were  investigated using accelerometers and by observing uniform generation of Faraday waves \citep{Harris2015,supplement}. Unless otherwise stated, the speaker cone was driven simultaneously at $f=80\,\text{Hz}$ and $f/2=40\,\text{Hz}$, with each driving signal fed into a separate voice coil. The acceleration of the bath was measured using two accelerometers mounted on the edges of the bath. For each accelerometer, the $\gamma_{40}$ and $\gamma_{80}$ acceleration amplitudes and phase shift $\Delta \phi$ were extracted through a nonlinear least squares fitting of the accelerometer signal to Eq.~\eqref{eq:acceleration}. The measured phase difference $\Delta\phi$ differed from the input value by a constant and has an uncertainty of $\pm3{^\circ}$. The measured values of $\gamma_f$ and $\gamma_{f/2}$ have an uncertainty of $\pm 0.05\,$g. Small droplets with radius less than $0.8\,\text{mm}$ were created by swiftly extracting a needle from the oil bath while larger droplets were created using a syringe. The droplets were imaged with a top-view camera at $4$ frames per second and with a side-view high-speed camera for a selection of cases at a typical frame rate of $4000$ frames per second. A light panel placed above the camera provided sufficient illumination for the top-view images while a bright LED source illuminated the droplets from the side for the high-speed videos. The size of each droplet was measured from the top-view images using a Hough circle transform. Further details are provided in the Supplemental Material \cite{supplement}.  
\begin{figure}
\centering
\includegraphics[width=\columnwidth]{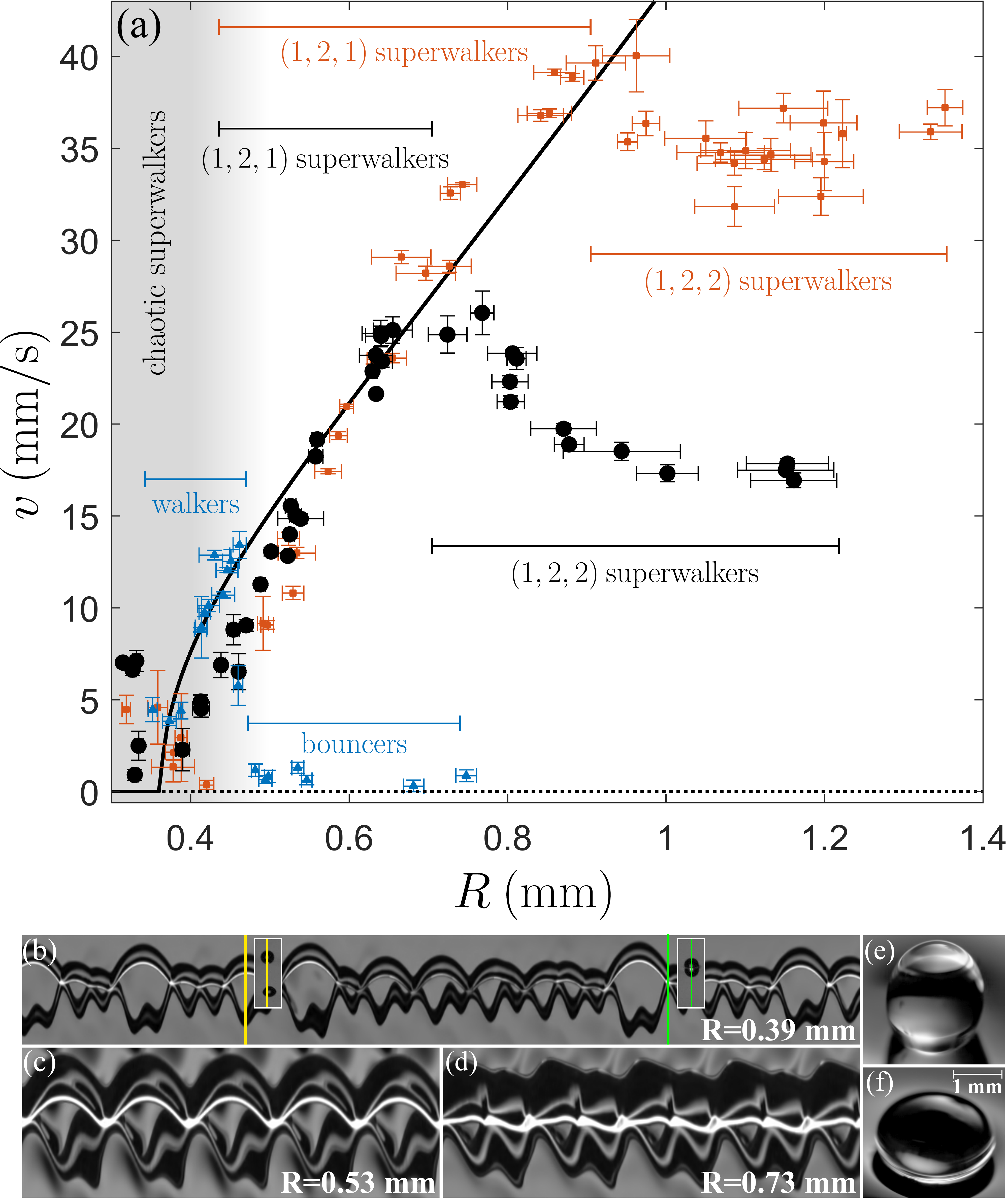}
\caption{Characteristics of a solitary superwalker. (a) Walking speed as a function of droplet radius for fixed values $\gamma_{80}=3.8\,\text{g}$ and $\Delta\phi=130^{\circ}$ and three different values of $\gamma_{40}$, specifically $\gamma_{40}=0\,\text{g}$ (blue markers), $\gamma_{40}=0.6\,\text{g}$ (black markers) and $\gamma_{40}=1.0\,\text{g}$ (red markers). Details of the error bars and the theory prediction (solid curve) are explained in the Supplemental Material \cite{supplement}. Three different bouncing behaviors are indicated for superwalkers: chaotic, \twoone{} and \twotwo{}. Vertical slice-time plots of droplets are shown for (b) chaotic, (c) \twoone{} and (d) \twotwo{} bouncing modes corresponding to the black markers at the radii indicated. These spatiotemporal images are generated by juxtaposing vertical
sections one pixel wide passing through the droplet's center of mass. Panels (e) and (f) show the two extremes of the shape deformations of a jumbo superwalker.}
\label{Fig: SS}
\end{figure}

\begin{figure*}
\centering
\includegraphics[width=2\columnwidth]{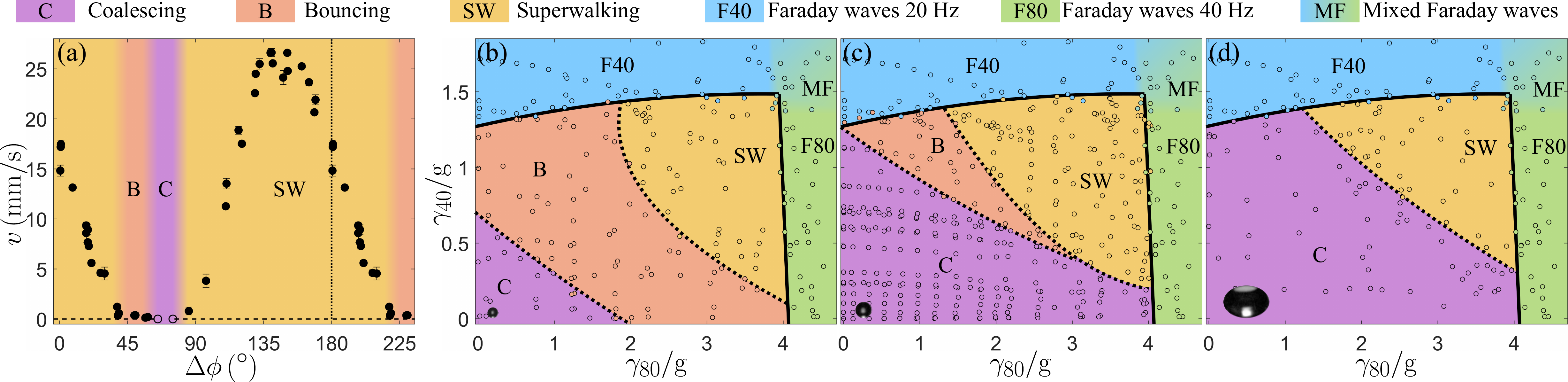}
\caption{(a) Droplet speed as a function of phase difference $\Delta\phi$ for a droplet of radius $R=0.83\pm0.03\,\text{mm}$ for $\gamma_{80}=3.8\,$g and $\gamma_{40}=0.6\,$g. The data to the right of the vertical dashed line is repeated. Different behaviors occurring in the $(\gamma_{80},\gamma_{40})/g$ parameter space for a fixed phase difference $\Delta\phi=130^{\circ}$ and three different droplet radii: (b) $R=0.6\pm0.05\,$mm, (c) $R=0.8\pm0.05\,$mm, and (d) $R=1.0\pm0.05\,$mm. Error bars are explained in the Supplemental Material \cite{supplement}.}
\label{Fig: PS}
\end{figure*}

Characteristics of solitary superwalkers are shown in Figs~\ref{Fig: SS} and~\ref{Fig: PS}. Figure~\ref{Fig: SS}(a) shows the speed of a droplet as a function of its radius for fixed values of $\gamma_{80}$ and $\Delta \phi$ and three values of $\gamma_{40}$, illustrating the significant size and speed increase possible for superwalkers. Three prominent types of walking are observed for two-frequency driving and are identified in Fig.~\ref{Fig: SS}(a). The smallest droplets, which are walkers for single-frequency driving, become chaotic superwalkers upon adding the subharmonic driving signal. These droplets bounce aperiodically, see Fig.~\ref{Fig: SS}(b), and walk unsteadily with significant fluctuations in their velocities. Similar irregular walking dynamics for two-frequency forcing at $80\,\text{Hz}$ and $64\,\text{Hz}$ has been observed previously \citep{PhysRevE.94.053112}.

Much larger droplets that would not be able to walk for single-frequency driving can walk with two frequencies $f$ and $f/2$.  They have constant velocity and typically have greater speeds than the fastest walkers.  Two different bouncing modes are realized for such superwalkers. We describe the vertical bouncing behavior of droplets driven by two frequencies using the generic notation \bmnot{} indicating that the droplet impacts the surface $n$ times during $m$ oscillation periods of the bath at frequency $f$, which equals $l$ oscillation periods of the bath at frequency $f/2$. For single frequency driving, the $l$ index is dropped. Superwalkers in the \twoone{} mode impact the bath once every two up-and-down motions of the bath as shown in Fig.~\ref{Fig: SS}(c) and their speed increases almost linearly with increasing size of the droplet. Surprisingly, despite the presence of $\gamma_{40}$ being essential to the existence of superwalkers, its magnitude only marginally affects the speed of \twoone{} superwalkers. This is consistent with the observations for (2,1) walkers, for which the walking speed is only weakly dependent on the driving amplitude at higher accelerations above the walking threshold \citep{Molacek2013DropsTheory}. 

Very large superwalkers bounce in a \twotwo{}{} mode, see Fig.~\ref{Fig: SS}(d). In contrast to the \twoone{} superwalkers, the speed of \twotwo{} superwalkers decreases with increasing droplet size. We attribute this behavior to the increased drag due to the prolonged contact time between the droplet and the bath. Superwalkers with radius $R \gtrsim 0.9\,$mm undergo significant internal deformation and do not seem to lift off from the surface. We refer to these as \textit{jumbo superwalkers}, see Figs~\ref{Fig: SS}(e,f) and Supplemental Video S1 \cite{supplement}. The frequency of the elliptical shape vibrations of the jumbo superwalkers is close to their bouncing frequency \citep{supplement,lamb1895hydrodynamics,prosperetti_1980}. Intriguingly, jumbo superwalkers cannot simply bounce without walking.

The solid curve in Fig.~\ref{Fig: SS}(a) is the speed-size relationship for hypothetical (2,1) walkers with single-frequency driving predicted from the Oza-Rosales-Bush stroboscopic model \citep{Oza2013,supplement}. The prediction is reasonably accurate, potentially because the bouncing mode and wavefield of (1,2,1) superwalkers are similar to (2,1) walkers. 
Since the stroboscopic model is only valid for walkers bouncing in a (2,1) mode, we do not expect it to be applicable for modeling the (1,2,2) superwalkers. However, recently more detailed numerical simulations \cite{Carlos2019} have also reproduced (1,2,1) superwalking behavior, and these may be able to describe (1,2,2) normal superwalker behavior as well.  Modelling of jumbo superwalkers will presumably require consideration of shape deformations.

The value of the phase difference $\Delta\phi$ between the two driving signals crucially affects the behavior of droplets. Figure~\ref{Fig: PS}(a) shows representative data for the speed of a superwalker as a function of $\Delta\phi$ for fixed droplet radius and acceleration amplitudes. We find that superwalkers only exist for a limited range of phase difference and outside this range they either coalesce (open markers) or bounce. For the parameters corresponding to Fig.~\ref{Fig: PS}(a), the droplets bounce in a \twotwo{} mode both in the bouncing and the superwalking regime. The maximum speed occurs in the vicinity of $\Delta\phi\approx140^{\circ}$, a value that does not appear to vary significantly with $\gamma_{80}$, $\gamma_{40}$, or droplet size.

Figures~\ref{Fig: PS}(b-d) show the different regimes observed in the $(\gamma_{80},\gamma_{40})$ parameter space for a fixed phase difference $\Delta \phi =130^\circ$ and three different droplet radii. Parametrically forcing a bath of liquid simultaneously at two different frequencies $f$ and $f/2$ may result in a Faraday instability with either $f/2$ or $f/4$ waves depending on the amplitudes of the two frequencies and the phase difference $\Delta\phi$ between them \citep{twofreqfaraday}. We find that driving the bath at $80\,$Hz and $40\,$Hz delays the onset of $20\,$Hz Faraday waves when the driving acceleration $\gamma_{80}$ is large. The onset of the $40\,$Hz Faraday waves is not significantly affected. For large $\gamma_{40}$ and $\gamma_{80}$, both $40\,$Hz and $20\,$Hz Faraday waves appear to be excited simultaneously. Below the Faraday threshold, we find coalescing (C), bouncing (B), and superwalking (SW) regions with the extent of each region dependent on droplet size. For a relatively small droplet, Fig.~\ref{Fig: PS}(b), the extent of the bouncing and superwalking regions is large. The bouncing region progressively decreases with an increase in droplet size, see Figs~\ref{Fig: PS}(b-d). For a larger droplet, Fig.~\ref{Fig: PS}(d), the bouncing region disappears and the droplet may either coalesce or walk. For even larger droplets, the superwalking region also vanishes. We also find that just above the $80\,$Hz-driving Faraday threshold, unlike walkers, superwalkers still walk steadily with their motion guided by the globally excited nonlinear Faraday waves. In the parameter regime where global Faraday waves are not excited, droplets always appear to trigger decaying $40\,$Hz Faraday waves, as illustrated by the similarity in wavelengths in Fig.~\ref{Fig: schematic}.

\begin{figure}
\centering
\includegraphics[width=\columnwidth]{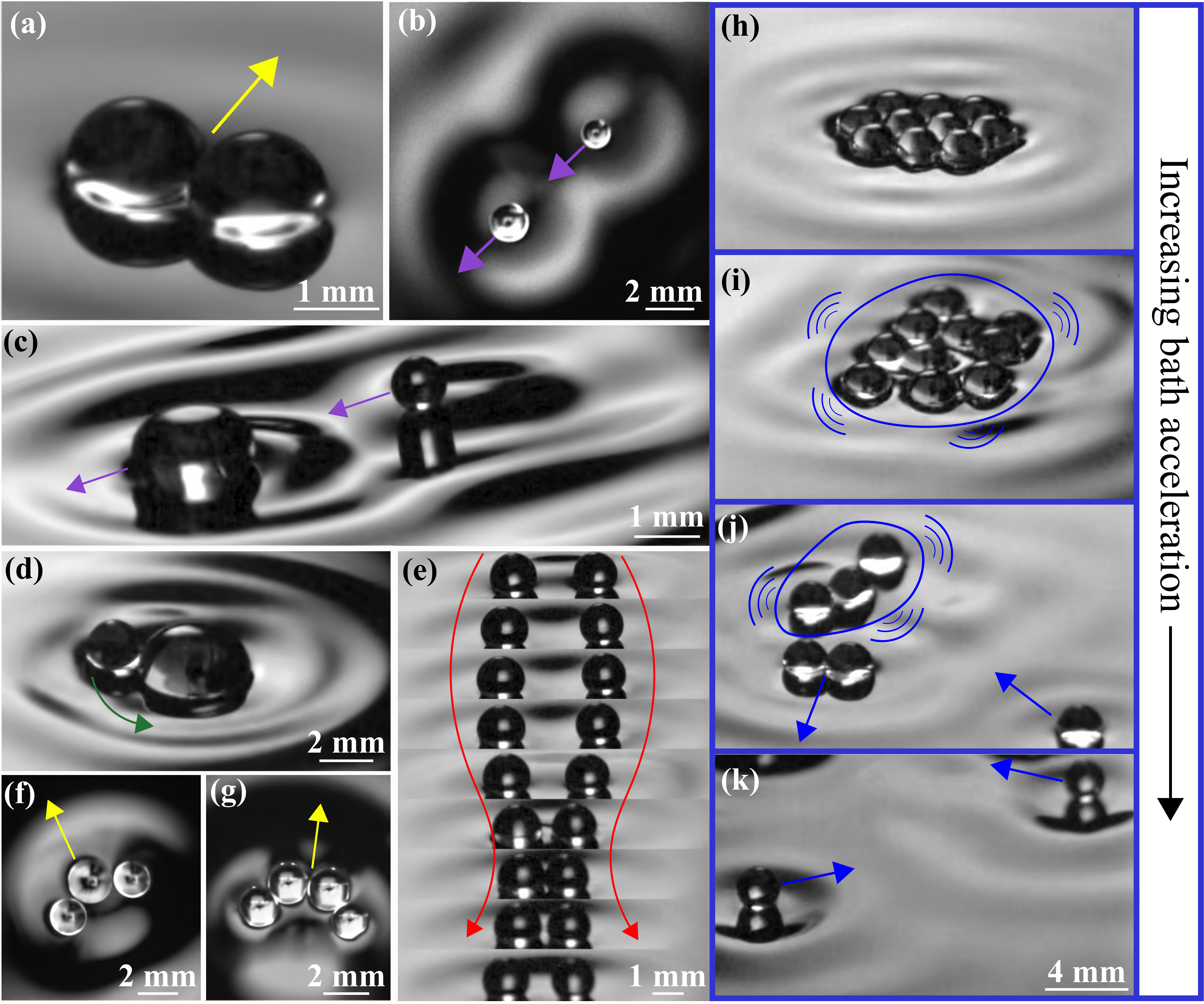}
\caption{A plethora of phenomena observed with superwalkers; see text for details.
}
\label{Fig: td}
\end{figure}

Superwalkers open up an extended parameter space to explore new phenomena with walking droplets, a selection of which are illustrated in Fig.~\ref{Fig: td}. Multiple superwalkers interacting with each other through their wave field form a variety of stationary and dynamic configurations. Two superwalkers can bind into a tight pair in which the droplets are separated only by a very thin air layer, see Fig.~\ref{Fig: td}(a) and Supplemental Video S2 \cite{supplement}. If such a pair of droplets have differing size they traverse a circular path, while a pair of identical droplets traverses a straight path. Similar states exist for staggered three-droplet and four-droplet configurations, see Figs~\ref{Fig: td}(f,g) and Supplemental Videos S3 and S4 \cite{supplement}.

We have observed another type of bound pair called chasers, see Figs~\ref{Fig: td}(b,c) and Supplemental Videos S5 and~S6 \cite{supplement}, which have previously been found numerically for identical droplets with single-frequency driving \cite{ValaniHOM,durey_milewski_2017} and experimentally in an effectively one-dimensional annular geometry  \cite{PhysRevE.92.041004}. In this state, two droplets walk one behind the other at a constant speed. For droplets of differing size, the larger droplet leads and drags the smaller one in its wake. Chasing pairs of superwalkers are robust and ubiquitous at high memory and the larger the size disparity between the two droplets, the more stably bound they are. Particularly disparate pairs can survive collisions with other droplets and even with the bath's walls.  Less commonly, we have observed droplet trains consisting of three chasing droplets in descending size order. We note that chasing pairs are different from ratcheting pairs of walkers \citep{Eddi2008,GaleanoRios2018}. Ratcheting motion typically occurs below the walking threshold and the pair travels slowly, while we find that chasers only appear at high memory and are an order of magnitude faster. 

 Two superwalkers can form a state reminiscent of promenading pairs of walkers, where the droplets walk in parallel with sideways oscillations \citep{Borghesi2014,PhysRevFluids.3.013604,twodroplets}. Promenading pairs of walkers remain physically separated at all times due to the wave barrier formed as they approach one another. In contrast, promenading pairs of superwalkers undergo droplet-droplet collisions, bouncing off one another, see Fig.~\ref{Fig: td}(e) and Supplemental Video S7. The center-of-mass of identical promenading superwalkers follows a straight path while that of even slightly mismatched superwalkers tends to follow a circular trajectory.

Two superwalkers may also form loosely bound orbiting pairs similar to walkers (see \citep{Protiere2006,PhysRevFluids.2.053601,PhysRevE.78.036204} for descriptions of ordinary walkers in this configuration).  A novel feature for superwalkers is that a size mismatch results in intermittent reversals of the orbiting direction. We also observe tightly bound orbiting pairs of mismatched superwalkers.  With an extreme size imbalance, giant droplets that coalesce with the bath in isolation can persist if accompanied by a smaller orbiting satellite droplet, see Fig.~\ref{Fig: td}(d) and Supplemental Video~S8 \cite{supplement}.

When many superwalkers are present, the inter-droplet interactions favor crystalline droplet configurations for relatively low driving accelerations. If the value of $\gamma_{40}$ is progressively increased while keeping $\gamma_{80}$ fixed, see Fig.~\ref{Fig: td}(h-k), the crystal initially begins to jiggle. Similar jiggling of a droplet crystal has been observed for single frequency driving on decreasing the frequency or increasing the number of droplets \citep{Lieber2007}. Increasing $\gamma_{40}$ further results into disintegration of the droplet crystal but droplets may still remain bounded in two and three droplet configurations. Ultimately, at highest $\gamma_{40}$, the droplets begin to superwalk at high speed, bouncing off each other elastically like billard balls. The observed dynamics are reminiscent of solid-liquid-gas phase transitions with the forcing amplitude acting as a temperature parameter, see Supplemental Video~S9 \cite{supplement}. This behaviour is robust with respect to interchanging the roles of $\gamma_{40}$ and $\gamma_{80}$, and are associated with crossing the phase boundary between bouncing (B) and superwalking (SW).

Another interesting dynamical phenomenon may be observed when the forcing frequencies are slightly detuned, for example with $80\,$Hz and $39.5\,$Hz driving, see Supplemental Video~S10 \cite{supplement}.  Here the droplets perform a \emph{stop-and-go motion} in which the droplets walk for a while, then stop abruptly, then walk again, and so on. Such motion arises because the slight detuning of the two frequencies causes the phase difference $\Delta \phi$ to evolve very slowly in time and so the droplet periodically cycles through the superwalking and bouncing regimes.  Even if a coalescence regime is encountered, coalescence may be avoided if such a regime is traversed quickly enough. The result is an effective discrete-time dynamical system that emerges out of an underlying continuous-time system.

In conclusion, we have introduced a new class of walking droplets, coined superwalkers, enabled by adding a subharmonic driving signal to a periodically driven walking-droplet system. This introductory study of superwalkers paves the way for a wealth of new phenomena and warrants further investigation.

\begin{acknowledgments}
We are grateful to D. Duke, K. Helmerson, S. Johnstone, S. Morton, and L. Suryawinata for technical assistance and J.C. Miller for useful conversations. We acknowledge financial support from an Australian Government Research
Training Program (RTP) Scholarship (R.V.) and the Australian Research Council via the Future Fellowship Project No.\ FT180100020 (T.S.).
\end{acknowledgments}

\bibliography{Superwalking_droplets}
\end{document}